\documentstyle[preprint,eqsecnum,aps]{revtex}
\begin{document}
\preprint{SNUTP 97-165}

\tightenlines
\draft

\title{Propagation of vacuum polarized photons in \\
topological black hole spacetimes}

\author{Rong-Gen Cai \footnote{Email address:cairg@ctp.snu.ac.kr}}
\address{Center for Theoretical Physics, Seoul National University,
Seoul 151-742, Korea}

\maketitle 

\begin{abstract}
The  one-loop effective action for QED in curved spacetime 
contains  equivalence principle violating interactions between the 
electromagnetic field and the spacetime curvature. These interactions 
lead  to  the dependence of photon velocity on the motion and 
polarization directions. In this paper we investigate the 
gravitational analogue to the electromagnetic birefringence phenomenon 
in the static and  radiating topological black hole backgrounds, 
respectively. For the    static topological black hole spacetimes,
the velocity shift of photons   is the same as the one in the 
Reissner-Nordstr\"om black holes. This  reflects  that the propagation 
of vacuum polarized photons is not sensitive
to the asymptotic behavior and topological structure of spacetimes.
For the massless topological black hole and BTZ black hole, the
light cone condition keeps  unchanged. In the radiating topological 
black hole backgrounds, the light cone condition is changed even 
for the radially directed photons. The velocity shifts  depend on the 
topological  structures. Due to the null fluid, the velocity shift of
 photons does no longer vanish at  the  apparent horizons as well as 
the event horizons.  But the ``polarization sum rule'' is still valid.

\end{abstract}
\pacs{PACS numbers: 03.50.De, 04.20.Gz, 04.60.+v \\
Keywords: Superluminal photon, Gravitational birefringence phenomenon, 
Topological black hole.}

\section{ Introduction}

The Hawking radiation \cite{Haw} of  black holes  is a  remarkable 
prediction of quantum field theory in curved spacetime. This has 
been leading to a quite active field now: black hole physics. Compared
to the Hawking radiation, the superluminal (``faster than light'')
photon propagation in  gravitational backgrounds, one of the
predictions of quantum field theory  in curved spacetime, is little 
to know.  This phenomenon was discovered first  by Drummond and Hathrell
\cite{Dru} in 1980. They calculated in QED the contribution to the photon
effective action from one-loop vacuum polarization on a general
gravitational background and used it to investigate  the correction 
to the local
propagation of photons in the optics approximation. They found that the 
quantum corrections introduce tidal gravitational forces on the photons
which in general alter the characteristics of propagation. They
investigated this phenomenon in the Schwarzschild spacetime, de Sitter
space, gravitational wave background and the Robertson-Walker
spacetimes.  The  photons may travel at speed greater than unity 
in certain  motion and polarization directions, 
except for the case of de Sitter space. Shortly after this discovery, 
Ohkuwa \cite{Ohku}  obtained the similar result for  the massless 
neutrino  propagating in the  Robertson-Walker spacetime.

It was not further studied until the work of Daniels and Shore
\cite{Dan1}, in which they generalized the analysis of Drummond and 
Hathrell to the Reissner-Nordstr\"om black hole spacetimes, and further
confirmed this phenomenon: for the radial motion photons the light cone
is unchanged and the photon velocity is still the unity; but for the
orbital photons the velocity may depend on the polarization directions.
This phenomenon is in fact the gravitational analogue to the
electromagnetic birefringence \cite{Alder,Dan1}. Due to the vacuum
polarization in QED, the photons exist for part of the time as a virtual 
$e^+e^-$ pair, thereby acquiring an effective size of the order of the
Compton wavelength of the electron ($\lambda_c=1/m_e$, $m_e$ is the mass 
of the electron). Thus the motion of  photons could be changed 
by the tidal effects of the spacetime curvature. But, ``faster than
light'' photons do not violate  the causality \cite{Dru,Dan1}, because 
the effective action contains the equivalence principle violating
interactions. More  recently, this phenomenon has been studied further 
in the Kerr black  hole background \cite{Dan2} and dilaton black hole 
spacetime \cite{Cho}. Due to the rotating feature of Kerr metric,
Daniels and Shore \cite{Dan2} found that the light cone is also changed
even for the radial motion photons. But the velocity 
shifts are always equal and opposite for two physical polarization
directions, and vanish at  the black hole event horizons. According to 
these observations, Shore \cite{Shore} proposed two theorems about 
the propagation of vacuum polarized photons: ``horizon theorem'' and
``polarization sum rule''. The ``horizon theorem'' states that the
velocity of radial motion photons remains equal to $c$ at the event 
horizons. This is satisfied by now for the de Sitter space,
Schwarzschild black hole, Reissner-Nordstr\"om black hole, Kerr black
hole and the dilaton black hole. This theorem 
seems to ensure that the geometric event horizon for black holes remains 
a true  horizon for real photon propagation in QED. The ``polarization 
sum rule'' states that the sum of the averaged velocity shifts for two 
physical polarization directions is proportional to the energy-momentum
tensor of matter. Hence  for Ricci flat spacetime, the sum over the two
physical polarizations of the velocity shift is zero. In \cite{Laf}
Lafrance and Myers investigated the propagation of  
photons  with the emphasis on the dispersive property of photon propagating
 in gravitational backgrounds and the validity of the equivalence
principle in QED.

The velocity shift of photons also takes place in non-gravitational 
backgrounds, for instance, in electromagnetic field \cite{Alder,Dan1},
Casimir-type regions with boundaries \cite{Schar,Bart}, and in the 
finite temperature and/or density backgrounds \cite{Lato}. Except the case 
in the Casimir-type regions, the velocity of photons is always less than 
$c$. In a flat spacetime, indeed the propagating velocity larger than $c$ 
may threat the causality. For the case of electromagnetic wave
traveling in vacuum between two parallel conducting plates, Ben-Menahem
\cite{Ben} showed that the wavefront still travels at exactly $c$. The
two-loop effect poses thus no threat to causality in QED. Contrary to 
the situation in flat spacetime, in curved spacetime the propagation of 
the so-called ``faster than light'' photons is possible and does not 
necessarily imply  that the causality must be violated. Usually, the 
establishment 
of a causal  paradox needs at least two conditions: spacelike motion
and Poincar\'e invariance, in a flat spacetime.  In curved spacetime the 
Poincar\'e invariance is lost and replacing it is the principle of 
equivalence. Just as mentioned above, due to the interacting terms
between the electromagnetic field and the spacetime curvature, the
principle of equivalence is lost in QED in curved spacetime. The
superluminal photons are therefore allowed in curved spacetime. This may
be an important feature of quantum field theory in curved spacetime.

Although the phenomenon of the superluminal photon propagation is
unmeasurable almost \cite{Dru}, it is still a quite interesting
phenomenon at least in principle. In order to better understand and 
to find its general feature,
it is necessary to investigate this phenomenon thoroughly. In the 
present paper we would like to extend previous investigations in three
directions. First, according to \cite{Dru}, the velocity of vacuum 
polarized photons remains unchanged in de Sitter space due to the 
isotropy of the spacetime. It is therefore interesting to see whether
the cosmological constant affects the velocity shift in other spacetimes
with a cosmological constant.  Secondly, we hope to know
whether the topological structure of backgrounds has effects on the 
propagation of vacuum polarized  photons. Thirdly, the ``polarization 
sum rule'' and `` horizon theorem'' have been checked for the static and 
stationary black hole backgrounds only. It is of some interest to
consider the phenomenon in the dynamical black hole spacetimes. For the
sake of generality, as the gravitational background we choose 
the solutions to Einstein-Maxwell equations with a  cosmological
constant. This theory contains  the static, spherically symmetric 
Reissner-Nordstr\"om-(anti-)de Sitter black hole solutions. The
spacetime is asymptotically anti-de Sitter or de Sitter one, 
depending  on the sign of the cosmological constant. The 
topologies of their  event horizons are both  a two-sphere $S^2$.
When the cosmological constant is negative, 
the theory has also the so-called topological black hole
solutions\cite{topo,Huang,Cai1,Amin,Mann,Bans,Brill,Van}. That
is, the two-dimensional hypersurface of event horizon may have zero 
curvature or negative constant curvature. Thus, the topology
of event horizon  is no longer the two-sphere. These topological black
hole solutions also exist in the dilaton gravity \cite{Cai2}.

The plan of this paper is as follows. In next section we introduce
the one-loop effective action in QED, and obtain the equation of vacuum 
polarized photons using the geometric optics approximation. In section 
3 we investigate the propagation of photons in the static topological 
black hole backgrounds. In section 4 we discuss the case in the
radiating  topological black holes. We present our conclusion 
in section 5.

 \section{ Effective action and equations of motion}

The one-loop effective action of QED in a general gravitational
background has been calculated in \cite{Dru}, and has been used 
already to discuss the propagation of photons in some gravitational 
fields, such  as the Schwarzschild black holes \cite{Dru}, 
Reissner-Nordstr\"om black holes \cite{Dan1}, Kerr black holes 
\cite{Dan2} and dilaton black holes \cite{Cho}.
The effective action has been also used to explain the production 
of primordial magnetic field in the early universe \cite{Maz}.
In order to compare
with these known results, we use  the following  one-loop effective 
action up to the same approximation,
\begin{eqnarray}
\label{action}
S=&-&\frac{1}{4}\int d^4x \sqrt{-g}F_{\mu\nu}F^{\mu\nu} \nonumber \\
  &-& \frac{1}{m^2_e}\int d^4x\sqrt{-g}(aRF_{\mu\nu}F^{\mu\nu} +
    bR_{\mu\nu}F^{\mu\sigma}F^{\nu}_{\ \sigma} +
	 c R_{\mu\nu\sigma\tau}F^{\mu\nu}F^{\sigma\tau} +
	 d \nabla _{\mu}F^{\mu\nu}\nabla _{\sigma}F^{\sigma}_{\ \nu}
	 \nonumber \\
 &+& \frac{1}{m^4_e}\int d^4x \sqrt{-g}\left [z(F_{\mu\nu}F^{\mu\nu})^2 
  +yF_{\mu\nu}F_{\sigma\tau}F^{\mu\sigma}F^{\nu\tau}\right ],
\end{eqnarray}
where $a$, $b$, $c$, $d$, $z$ and $y$ are constants. 
For the QED correction, these coefficients  are \cite{Dru,Maz}
\begin{eqnarray}
\label{coef}
&& a=-\frac{5}{720}\frac{\alpha}{\pi}, \ \
   b=\frac{26}{720}\frac{\alpha}{\pi}, \ \ 
   c=-\frac{2}{720}\frac{\alpha}{\pi}, \nonumber \\
&& d=-\frac{1}{30}\frac{\alpha}{\pi}, \ \ 
   z=-\frac{5}{180}\alpha ^2, \ \
	y=\frac{14}{180}\alpha ^2.
\end{eqnarray}	
Here  $\alpha $ is the fine structure constant.
It is clear that the first line in (\ref{action}) is the usual action
of Maxwell field. The first three terms in the second line reveal the
influences of the curvature, and the fourth survives even in flat
spacetime and represents off-mass-shell effects in the vacuum
polarization. The third line is the  Euler-Heisenberg terms. 
 This action (\ref{action}) is valid in the approximation of weak 
curvature and low frequency photons. Therefore the higher powers of the 
curvature  tensor and extra covariant derivatives in the interacting 
terms can be  neglected. Varying the action (\ref{action}) yields 
the  equation of motion for the  electromagnetic field,
\begin{eqnarray}
\label{eq}
\nabla _{\mu}F^{\mu\nu} &+& \frac{1}{m^2_e} [ 4a \nabla
            _{\mu}(RF^{\mu\nu})+2b \nabla _{\mu}(R^{\mu}_{\ \sigma}
	F^{\sigma\nu}-R^{\nu}_{\ \sigma}F^{\sigma\mu}) 
	\nonumber \\
	&+& 4c \nabla _{\mu}(R^{\mu\nu}_{\ \
	\sigma\tau}F^{\sigma\tau}) +2d (\nabla ^2 \nabla
	_{\sigma}F^{\sigma\nu}-\nabla_{\mu}\nabla ^{\nu}\nabla
	_{\sigma}F^{\sigma\mu})]\nonumber \\
	&-&\frac{1}{m^4_e}\left
	[8z(F^{\sigma\tau}F_{\sigma\tau}\nabla _{\mu}F^{\mu\nu}
	+ 2F^{\mu\nu}F_{\sigma\tau}\nabla_{\mu}F^{\sigma\tau})
	\right. \nonumber \\
	&+&\left. 8y (F^{\nu\tau}F_{\sigma\tau}\nabla _{\mu}
	F^{\mu\sigma}+F^{\mu\sigma}F_{\sigma\tau}\nabla_{\mu}
	F^{\nu\tau}+F^{\mu\sigma}F^{\nu\tau}\nabla_{\mu}
	F^{\sigma\tau})\right ]=0.
\end{eqnarray}				
In order to study the propagation of photons, we expand the field 
strength $F^{\mu\nu}$ using background field method,
\begin{equation}
\label{expansion}
F^{\mu\nu}=\bar{F}_{\mu\nu}+\hat{f}_{\mu\nu},
\end{equation}
where $\bar{F}_{\mu\nu}$ is the background electromagnetic field.
Substituting (\ref{expansion}) into (\ref{eq}) and linearizing the
equation in $\hat{f}_{\mu\nu}$, one can get the equation describing 
the propagation of photons,
\begin{eqnarray}
\label{eqph}
\nabla _{\mu}\hat{f}^{\mu\nu} &+& \frac{1}{m^2_e}[2b 
         R^{\mu}_{\ \sigma}
          \nabla _{\mu}\hat{f}^{\sigma\nu} +4c R^{\mu\nu}_{\ \
	 \sigma\tau}\nabla _{\mu}\hat{f}^{\sigma\tau}]
	 \nonumber \\
	 &-&\frac{1}{m^4_e}\left[16z
	 \bar{F}^{\mu\nu}\bar{F}_{\sigma\tau}\nabla_{\mu}
	 \hat{f}^{\sigma\tau}+ 8y \left(
	 \bar{F}^{\mu\sigma}\bar{F}_{\sigma\tau}\nabla_{\mu}
	 \hat{f}^{\nu\tau} +\bar{F}^{\mu\sigma} 
	 \bar{F}^{\nu\tau}\nabla _{\mu}\hat{f}_{\sigma\tau}\right)
	 \right]=0.
\end{eqnarray}			 
Here some  approximations have been used to derive
Eq.~(\ref{eqph}). The quantum corrections are retained up to the first 
order of $\alpha$ from the background curvature and
electromagnetic field. The typical variations of the background
electromagnetic field and gravitational field, characterized by the
scale $L$, are assumed to be much small than the one of the photons.
That is, $L>>\lambda$, here $\lambda$ is the wavelength of photons. Thus 
some  derivative terms, such as $\nabla_{\sigma} \bar{F}_{\mu\nu}$, 
$\nabla_{\sigma} R_{\mu\nu}$, etc. can
be omitted. Without the one-loop correction, we have
$\nabla_{\mu}F^{\mu\nu}=0$. Hence $\nabla_{\mu}\hat{f}^{\mu\nu}$ gives
at least the first order correction of $\alpha$. Such terms in
interacting terms can also be
neglected. Other assumption in Eq.~(\ref{eqph}) is that the wavelength 
of photons is much larger than that of the Compton wavelength:
$\lambda>>\lambda_c$. The details of these approximations can be found
in \cite{Dan1} and \cite{Cho}.

In order to investigate the behavior of propagation of vacuum 
polarized photons, a  simple method is to employ the geometric 
optics approximation \cite{Dru}. In this method, one can write
\begin{equation}
\label{optics}
\hat{f}_{\mu\nu}=f_{\mu\nu}e^{i\theta},
\end{equation}
where $f_{\mu\nu}$ is a slowly varying amplitude and $\theta$ the
rapidly varying phase. 
The wave vector is $k_{\mu}=\nabla _{\mu}\theta$. In the quantum
mechanics, it can be regarded as the momentum of photons. Thus from
the electromagnetic Bianchi identity 
\begin{equation}
\nabla_{\lambda}F_{\mu\nu} +\nabla
_{\mu}F_{\nu\lambda}+\nabla_{\nu}F_{\lambda \mu}=0,
\end{equation}
one has 
\begin{equation}
k_{\lambda}f_{\mu\nu}+k_{\mu}f_{\nu\lambda}+k_{\nu}f_{\lambda\mu}=0.
\end{equation}
Further  we can write down 
\begin{equation}
\label{polar}
f_{\mu\nu}=k_{\mu}a_{\nu}-k_{\nu}a_{\mu},
\end{equation}
where  $a_{\nu}$ can be interpreted as the polarization vector of 
photons and satisfies  $k_{\mu}a^{\mu}=0$. Substituting (\ref{optics}) 
into (\ref{eqph}) and using the relation (\ref{polar}), we arrive at
\begin{eqnarray}
\label{eqph2}
k^2a^{\nu} &+& \frac{1}{m^2_e}[2b R^{\mu}_{\ \sigma}
        k_{\mu}(k^{\sigma}a^{\nu}-k_{\nu}a^{\sigma})
	 +8c R^{\mu\nu}_{\ \
	 \sigma\tau}k_{\mu}k^{\sigma}a^{\tau}]
	 \nonumber \\
	 &-&\frac{1}{m^4_e}\left[32z
	 \bar{F}^{\mu\nu}\bar{F}_{\sigma\tau}k_{\mu}
	 k^{\sigma}a^{\tau} + 8y \left(
	 \bar{F}^{\mu\sigma}\bar{F}_{\sigma\tau}k_{\mu}(k^{\nu}a^{\tau}-
	 k^{\tau}a^{\nu})-\bar{F}^{\mu\sigma} 
	 \bar{F}^{\nu\tau}k _{\mu}k_{\tau}a_{\sigma}
	 \right)  \right]\nonumber \\
	 &=&0.
\end{eqnarray}

To study the propagation of photons in  curved spacetimes, it is  
 convenient to introduce the  orthonormal frame by using the 
vierbeins 
defined as $g_{\mu\nu}=\eta_{ab}e^{a}_{\ \mu}e^{b}_{\ \nu}$, where
$\eta_{ab}$ is the metric of Minkowski space. In the orthonormal
frames the equation (\ref{eqph2}) becomes 
\begin{eqnarray}
\label{eqph3}
k^2a^{b} &+& \frac{1}{m^2_e}[2b R^{a}_{\ c}
        k_{a}(k^{c}a^{b}-k^{b}a^{c})
      +8c R^{ab}_{\ \ cd}k_{a}k^{c}a^{d}]
       \nonumber \\
 &-&\frac{1}{m^4_e}\left[32z \bar{F}^{ab}\bar{F}_{cd}k_{a}
 k^{c}a^{d} + 8y \left( \bar{F}^{ac}\bar{F}_{cd}k_{a}(k^{b}a^{d}-
 k^{d}a^{b})-\bar{F}^{ac} \bar{F}^{bd}k _{a}k_{d}a_{c}
 \right)  \right]\nonumber \\
 &=&0.
\end{eqnarray}
In the next section, we will use this equation to investigate the
propagation of photons in the topological black hole backgrounds.

\section{Photons in  static topological black hole spacetimes}

Consider the Einstein-Maxwell equations with a cosmological constant,
$\Lambda$,
\begin{eqnarray}
\label{eqs1}
&& R_{\mu\nu}-\frac{1}{2}g_{\mu\nu}R +\Lambda g_{\mu\nu}=8\pi 
      T_{\mu\nu}^{\rm em},  \\
\label{eqs2}
&& \nabla_{\mu}F^{\mu\nu}=0,
\end{eqnarray}
where $T_{\mu\nu}^{\rm em}$ is the energy-momentum tensor of 
electromagnetic field,
\begin{equation}
\label{emtensor}
T_{\mu\nu}^{\rm em}=F_{\mu\sigma}F_{\nu}^{\ \sigma}
          -\frac{1}{4} g_{\mu\nu}F^2.
\end{equation}
In Eqs.~(\ref{eqs1}) and (\ref{eqs2}), there are well-known static,
spherically symmetric Reissner-Nordstr\"om-(anti-)de Sitter solutions,
\begin{eqnarray}
\label{rn}
&& ds^2=-f(r)dt^2 + f^{-1}(r)dr^2 +r^2d\theta ^2 +r^2\sin^2\theta
         d\phi^2,\\
&& F_{tr}=\frac{Q}{4\pi r^2},
\end{eqnarray}
 where
 \begin{equation}
\label{rn1}
f(r)=1-\frac{2M}{r}+\frac{Q^2}{4\pi r^2}-\frac{1}{3}\Lambda r^2.
\end{equation}
Here in order to compare with the case of Reissner-Nordstr\"om black
holes, we have used the same units as those in \cite{Dan1}. In equation 
(\ref{rn1}) the constants $M$ and $Q$ are the mass and electric charge 
of black holes, respectively. As $\Lambda <0$, the solution (\ref{rn})
describes an  asymptotically anti-de Sitter spacetime. Usually, the
equation, $f(r)=0$, has two positive roots. The large one $r_+$ is the
location of outer event horizon of black holes and the small $r_-$ 
the inner horizon
of black holes. As $\Lambda >0$, the solution (\ref{rn}) is the
Reissner-Nordstr\"om-de Sitter spacetime. In this case, the equation 
$f(r)=0$ may have three positive roots. The largest one $r_c$ is the 
cosmological horizon, the intermediate is the outer event 
horizon of the black
holes and the smallest is the inner horizon.

Although the asymptotic 
behaviors are quite different for the Reissner-Nordstr\"om-anti-de
Sitter black holes and Reissner-Nordstr\"om-de Sitter black holes, the
topological structures of their event horizons are same. They are both
the two-sphere $S^2$, as that of Reissner-Nordstr\"om black holes.
In recent years, many authors have found that, when the cosmological
constant is negative, the two-dimensional hypersurface of event horizon 
may have zero or negative constant curvature
\cite{topo,Huang,Cai1,Amin,Mann,Bans,Brill,Van,Cai2}. The topology of
event horizon of black holes is no longer the two-sphere $S^2$. For
instance, in the Einstein-Maxwell equations (\ref{eqs1}) and
(\ref{eqs2}), we have exact static solutions,
\begin{eqnarray}
\label{topo1}
 ds^2 &=& -\left (-\frac{2M}{r}+\frac{Q^2}{4\pi r^2}-\frac{1}{3}
        \Lambda r^2 
      \right )dt^2 +\left (-\frac{2M}{r}+\frac{Q^2}{4\pi r^2}
		-\frac{1}{3}\Lambda r^2\right )^{-1}dr^2 
    +r^2(d\theta ^2 + \theta ^2 d\phi^2 ), \\
\label{topo2}		
 ds^2 &=& -\left (-1-\frac{2M}{r}+\frac{Q^2}{4\pi r^2}-\frac{1}{3}
        \Lambda r^2 
      \right )dt^2 +\left (-1-\frac{2M}{r}+\frac{Q^2}{4\pi r^2}
		-\frac{1}{3}\Lambda r^2\right )^{-1}dr^2 \nonumber \\
		&& + r^2(d\theta ^2 + \sinh^2 \theta d\phi^2 ).
\end{eqnarray}		
Obviously, when $\Lambda <0$, both  the solutions (\ref{topo1}) and
(\ref{topo2}) are of the black hole structure in  certain parameter
regime. But, the scalar curvature of event horizon surface is zero for 
the solution (\ref{topo1}), and minus one for the solution (\ref{topo2}).
By appropriately identifying the coordinates $\theta$ and $\phi$, one
may obtain the topological black hole solutions whose event horizons 
possess different  topological structures. Combining the
Reissner-Nordstr\"om-(anti-)de Sitter solution (\ref{rn}) with 
the solutions (\ref{topo1}) and (\ref{topo2}), we  have the unified  
form
\begin{equation}
\label{topo}
ds^2=-f(r)dt^2 +f^{-1}(r)dr^2 +r^2 [d\theta ^2 +
       h_k^2(\theta )d\phi^2],
\end{equation}
where 
\begin{equation}
f(r)=k-\frac{2M}{r}+\frac{Q^2}{4\pi r^2} -\frac{1}{3}\Lambda r^2,
\end{equation}
and
\begin{equation}
h_k(\theta)= \left \{
\begin{array}{ll}
\sin \theta, & {\rm for}\ \ k=1 \\
\theta , &{\rm for}\ \ k=0   \\
\sinh \theta, & {\rm for} \ \ k=-1.
\end{array} \right.
\end{equation}
For the solution (\ref{topo}) the appropriate basis 1-forms are
\begin{equation}
\label{basis1}
e^0=\sqrt{f}dt, \ \ e^1=(\sqrt{f})^{-1}dr, \ \ e^2=rd\theta, \ \
   e^3=r h_k  d\phi.
\end{equation}		 
Through a straightforward calculation, in the orthonormal frame 
the nonvanishing components of curvature tensor are
\begin{eqnarray}
\label{riecomp1}
&& R_{0101}=\frac{f''}{2}\equiv A(r),     \ \
              R_{0202}=R_{0303}=\frac{f'}{2r}\equiv B(r),
               \nonumber \\
&& R_{1212}=R_{1313}=-\frac{f'}{2r}=-B(r),\ \
           R_{2323}=\frac{k-f}{r^2}\equiv D(r).
\end{eqnarray}
where a prime stands for the derivative with respect to $r$. Introducing
the notation $U^{01}_{ab}\equiv \delta^0_a\delta^1_b-
\delta ^1_a\delta^0_b$, etc., the Riemann  tensor can be
 expressed as 
 \begin{equation}
 \label{rie1}
 R_{abcd}=AU^{01}_{ab}U^{01}_{cd} +B (U^{02}_{ab}U^{02}_{cd}
       +U^{03}_{ab}U^{03}_{cd})-B(U^{12}_{ab}U^{12}_{cd}+
		 U^{13}_{ab}U^{13}_{cd})+D U^{23}_{ab}U^{23}_{cd},
\end{equation}
and the background electric field 
\begin{equation}
\label{field1}
\bar{F}_{ab}=\frac{Q}{4\pi r^2}U^{01}_{ab}.
\end{equation}
In order to solve the equation of motion of photons (\ref{eqph3}),
following \cite{Dan2}, we introduce some linearly independent 
combinations of momentum   components
\begin{equation}
l_b=k^aU^{01}_{ab}, \ \ m_b=k^aU^{02}_{ab},  \ \ n_b=k^aU^{03}_{ab},
\end{equation}
 and some dependent combinations
\begin{equation}
p_b=k^aU^{12}_{ab}, \ \ q_b=k^aU^{13}_{ab}, \ \ r_b=k^aU^{23}_{ab}.
\end{equation}
Using $l_a$, $m_a$, and $n_a$ to contract (\ref{eqph3}), respectively,
with the help of Eqs.~(\ref{rie1}) and (\ref{field1}), we have 
\begin{eqnarray}
\label{eqph5}
k^2 (a \cdot v) &+& \frac{2b}{m^2_e}(a \cdot v)
           [Al^2 +B(m^2 +n^2 -p^2 -q^2)  +Dr^2] \nonumber \\
	&+& \frac{8c}{m^2_e}[A (a \cdot l)(l\cdot v) + 
	B (m \cdot  v)(a \cdot m) +
	B (n\cdot v)(a\cdot n) \nonumber \\
	&& -B(a\cdot p)(v\cdot p) 
	-B(a\cdot q)(v\cdot q) +D(r\cdot a)(r\cdot v)] 
	\nonumber \\
	&-&\frac{1}{m^4_e}\left (\frac{Q}{4\pi r^2}\right)^2 
	\left \{32z(a\cdot l)(l\cdot v)+8y[l^2(a\cdot v) 
       +(l\cdot a)(l\cdot v)]\right \}=0
\end{eqnarray}
where $v=l$, $m$, $n$, respectively. Now we discuss the radial and orbital 
photon motions, respectively.

(i). {\it Radial photon motion}. In this case, we have $k^2=k^3=0$.
Hence,
\begin{eqnarray}
\label{vector1}
&& l^a=(k^1,k^0,0,0),\ \ m^a=(0,0,k^0,0), \ \ n^a=(0,0,0,k^0),\\
&& p_b=\frac{k^1}{k^0}m_b, \ \ q_b=\frac{k^1}{k^0}n_b, \ \ r_b=0.
\end{eqnarray}
For the longitudinal polarization $(a\cdot l)$, it follows from
Eq.~(\ref{eqph5})
\begin{equation}
\left [k^2 +\frac{2b}{m^2_e}(A+2B)l^2 +\frac{8c}{m^2_e}Al^2 
      -\frac{1}{m^4_e} \left(\frac{Q}{4\pi r^2}\right)^2(32z+16y)l^2 
		\right](a\cdot l)=0.
\end{equation}
Note that  $l^2=-k^2$ from (\ref{vector1}). It is easy to see that the
light cone condition $(k^2=0)$ is unchanged and the
velocity $|k^0/k^1|_r=1$ for the
longitudinal polarization of radial photons. For the physical 
transverse polarizations $(a\cdot m)$ and $(a\cdot n)$, simplifying 
(\ref{eqph5}) one  can get
\begin{equation}
k^2+\frac{2b}{m^2_e}(A+2B)l^2- \frac{8c}{m^2_e}Bk^2 -\frac{8y}{m^4_e}
    \left(\frac{Q}{4\pi r^2}\right)^2 l^2=0,
\end{equation}
from which again we have  $k^2=-l^2=0$. That is, 
\begin{equation}
\left|\frac{k^0}{k^1}\right|_{\rm \theta,\ \phi}=1.
\end{equation}
Therefore, the velocity of photons is still $c$ for the radially 
directed photons in the static topological black hole background. 
This is  independent of the topological structure of spacetimes.

(ii). {\it Orbital photon motion}. In this case we set $k^1=k^2=0$ 
and  consider photon propagation in the orbital ($\phi$)-direction. 
Thus we have 
\begin{eqnarray}
&& l^a=(0,k^0,0,0), \  \ m^a=(0,0,k^0,0), \ \ n^a=(k^3, 0, 0, k^0), \\
&& p_b=0,\ \ q_b=-\frac{k^3}{k^0}l_b, \ \ r_b=-\frac{k^3}{k^0}m_b.
\end{eqnarray}
For the unphysical longitudinal polarization $(a\cdot n)$, from
(\ref{eqph5}) one has
\begin{eqnarray}
\left \{k^2 \right. &+& \frac{2b}{m^2_e}[A(k^0)^2 +
      2 B(k^0)^2- 2B(k^3)^2+D(k^3)^2 ] \nonumber \\
   &+& \left. \frac{8c}{m^2_e}B[(k^0)^2-(k^3)^2]
	 -\frac{8y}{m^4_e}\left(\frac{Q}{4\pi r^2}\right)^2
	(k^0)^2 \right \}(a\cdot n)=0.
\end{eqnarray}
The photon velocity is, up to the first order correction,
\begin{equation}
\label{phivel1}
\left |\frac{k^0}{k^3}\right |_{\rm \phi}
				 =1+\frac{b}{m^2_e}(A+D)-\frac{4y}{m^4_e}
             \left(\frac{Q}{4\pi r^2}\right)^2.
\end{equation}
For the physical radial polarization $(a\cdot l)$,
\begin{eqnarray}
\left \{ \right. k^2 &+& \frac{2b}{m^2_e}[A(k^0)^2 +2B(k^0)^2-
         2B(k^3)^2 +D(k^3)^2 ] 
       +\frac{8c}{m^2_e}[A(k^0)^2 -B (k^3)^2] \nonumber \\
		 &-&\left. \frac{1}{m^4_e}\left(\frac{Q}{4\pi r^2}\right)^2(32z+
		 16y)(k^0)^2 \right \}(a\cdot l)=0,
\end{eqnarray}
from which we obtain the velocity of photons 
\begin{equation}
\label{rvel1}
\left |\frac{k^0}{k^3}\right|_{\rm r}=1+\frac{b}{m^2_e}(A+D)
       +\frac{4c}{m^2_e}(A-B)-\frac{16z+8y}{m^4_e}\left (\frac{Q}{4\pi
		 r^2}\right)^2 .
\end{equation}		 
For the other physical polarization $(a\cdot m)$, we have 
\begin{eqnarray}
\left \{ \right. k^2 + \frac{2b}{m^2_e}[A(k^0)^2  &+&
             2B(k^0)^2-2B(k^3)^2 +D(k^3)^2 ] 
       +\frac{8c}{m^2_e}[B(k^0)^2 +D(k^3)^2] \nonumber \\
		 &-& \left. \frac{8y}{m^4_e}\left(\frac{Q}{4\pi
		 r^2}\right)^2(k^0)^2 
		 \right \}(a\cdot m)=0.
\end{eqnarray}
In this polarization direction, the velocity of photons is
\begin{equation}
\label{thetavel1}
\left|\frac{k^0}{k^3}\right|_{\rm \theta}=
        1+\frac{b}{m^2_e}(A+D)+\frac{4c}{m^2_e}(B+D)
		  -\frac{4y}{m^4_e}\left(
		  \frac{Q}{4\pi r^2}\right)^2.
\end{equation}
From Eq.~(\ref{riecomp1}), a straightforward calculation gives 
\begin{eqnarray}
&& A=-\frac{2M}{r^3}+\frac{3Q^2}{4\pi r^4}-\frac{1}{3}\Lambda, 
      \nonumber   \\
&& B=\frac{M}{r^3}-\frac{Q^2}{4\pi r^4}-\frac{1}{3}\Lambda, 
      \nonumber \\
&& D=\frac{2M}{r^3} -\frac{Q^2}{4\pi r^4} +\frac{1}{3}\Lambda.
\end{eqnarray}
Substituting the above into Eqs.~(\ref{rvel1}), (\ref{thetavel1}) and
(\ref{phivel1}), we obtain
\begin{eqnarray}
&& \left|\frac{k^0}{k^3}\right|_{\rm \phi}=1+\frac{b}{m^2_e}\frac 
        {2Q^2}{4\pi
          r^4}-\frac{4y}{m^4_e}\left(\frac{Q}{4\pi r^2}\right)^2, \\
&& \left |\frac{k^0}{k^3}\right |_{\rm
          r}=1+\frac{b}{m^2_e}\frac{2Q^2}{4\pi r^4}
	 +\frac{4c}{m^2_e}\left(-\frac{3M}{r^3}+\frac{4Q^2}{4\pi
	 r^4}\right)
	 -\frac{16z+8y}{m^4_e}
        \left( \frac{Q}{4\pi r^2}\right )^2, \\
&& \left|\frac{k^0}{k^3}\right |_{\rm
          \theta}=1+\frac{b}{m^2_e}\frac{2Q^2}{4\pi r^4}
	 +\frac{4c}{m^2_e}\left(\frac{3M}{r^3}-\frac{2Q^2}{4\pi
	 r^4}\right) -\frac{4y}{m^4_e}\left(\frac{Q}{4\pi
	 r^2}\right)^2.
\end{eqnarray}			
For QED correction (\ref{coef}), they reduce to
\begin{eqnarray}
\label{shift1}
&& \left |\frac{k^0}{k^3}\right |_{\phi}=
        1+\frac{13}{45}\frac{\alpha}{m^2_e}\left(\frac{Q}{4\pi
	 r^2}\right)^2-\frac{14}{45}\frac{\alpha ^2}{m^4_e}
	 \left(\frac{Q}{4\pi r^2}\right)^2, \\
\label{shift2}
&& \left |\frac{k^0}{k^3}\right|_{r}=1+\frac{1}{30\pi}\frac{\alpha}
        {m^2_e}\frac{M}{r^3}+\frac{1}{9}\frac{\alpha}{m^2_e}
  \left(\frac{Q}{4\pi r^2}\right)^2-\frac{8}{45}\frac{\alpha
  ^2}{m^4_e}\left(\frac{Q}{4\pi r^2}\right)^2, \\
&&\left |\frac{k^0}{k^3}\right|_{\theta}=1-\frac{1}{30\pi }\frac{\alpha}
   {m^2_e}\frac{M}{r^3}+\frac{17}{45}\frac{\alpha}{m^2_e}
  \left(\frac{Q}{4\pi r^2}\right)^2-\frac{14}{45}\frac{\alpha
  ^2}{m^4_e}\left(\frac{Q}{4\pi r^2}\right )^2.
\label{shift3}
\end{eqnarray}			  
Comparing  with the results in 
\cite{Dan1,Cho}, we  find that the velocities of 
photon propagating in the topological black hole spacetimes are totally 
the same as  those in the Reissner-Nordstr\"om black hole backgrounds.
This reveals at least three noticeable  features. The first is that the
cosmological constant does not enter the expressions of velocity
 and hence the asymptotic behavior of spacetimes dose not  affect
the propagation of photons. In fact, this should be expected. When $M=Q=0$, 
the solution (\ref{topo}) reduces to the (anti-)de Sitter space. Therefore 
these results (\ref{shift1})-(\ref{shift3}) should go back to those in 
(anti-) de Sitter space. In addition,  when $Q=0$, these expressions are
 the same as those in the Schwarzschild black hole background \cite{Dru}. 
It is clear from (\ref{shift2}) and (\ref{shift3}) that the velocity shifts 
of the two physical polarizations are  equal and opposite. The averaged 
velocity shift is thus zero for neutral topological black holes. That
is, the cosmological constant does not change the result in the Ricci 
flat spacetimes.  The second point is that the
parameter $k$ representing different topological structures of 
spacetimes also 
does not enter the expressions of velocity. This reflects that the
propagation of photons is not sensitive to the topological structures 
of spacetimes. Finally, it is worth noticing that, when $\Lambda <0$, 
there is  still black hole structure  in Eq.~(\ref{topo}) even as 
$M=Q=0$,
\begin{equation}
ds^2=-(-1-\Lambda r^2/3)dt^2 +(-1-\Lambda r^2/3)^{-1}dr^2 +r^2(d\theta
       ^2 +\sinh^2\theta \phi^2).
\end{equation}		 
This so-called massless topological black hole \cite{Amin} has an event 
horizon at $r_+=\sqrt{3/|\Lambda|}$. In this case, the light cone keeps
unchanged and the velocity of photons is $c$ for any motion 
 and polarization directions. This massless topological black hole can
 be constructed by identifying certain points in a four dimensional
 anti-de Sitter space. This is reminiscent of the
 Ba\~nados-Teitelboim-Zanelli (BTZ)
 black hole in three dimensions \cite{BTZ}. The metric of BTZ black
 holes can be written as
 \begin{equation}
 \label{btz}
 ds^2=-N^2(r)dt^2 +N^{-2}(r)dr^2 +r^2 [ N^{\phi}(r)dt+d\phi ]^2,
 \end{equation}
 where
 \begin{equation}
 N^2(r)=-M -\Lambda r^2 + \frac{J^2}{4r^2}, \ \
        N^{\phi}(r)=-\frac{J}{2r^2},
\end{equation}
$M$ and $J$ are two integration constants and represent 
the mass and angular momentum of the holes, respectively. When
$J<M/\sqrt{|\Lambda|}$, the BTZ black holes have two  horizons
determined by the equation $N^2(r)=0$. The most remarkable feature of
the black hole is that this is a negative constant curvature 
solution and can be
constructed by identifying certain points in a three-dimensional anti-de
Sitter space. Therefore the Riemann tensor of the solution can be
expressed as \cite{BTZ}
\begin{equation}
\label{riebtz}
R_{abcd}=\Lambda (\eta _{ac}\eta_{bd}-\eta_{bc}\eta_{ad}).
\end{equation}

We now consider the propagation of vacuum polarized photons in the BTZ 
black hole background (\ref{btz}).
In this case,  the one-loop effective action in three
dimensions is different from the one in four dimensions (\ref{action}).
But we assume that this action (\ref{action}) is also valid in three 
dimensions, of course, the coefficients must not be  those in
Eq.~(\ref{coef}). Note that we can set $ d=z=y=0$ in this case to the
first order correction.
Using the equations (\ref{eq}) and (\ref{riebtz}), 
 we can  obtain
\begin{equation}
\beta \nabla_{\mu}F^{\mu\nu}=0,
\end{equation}
where $\beta $ is a constant.
Thus, this situation  is the same as that of the (anti-) de 
Sitter space in
four dimensions \cite{Dru}. Therefore the light cone is also 
unchanged in the BTZ  black hole background. 
Note that the BTZ
black hole is the analogue of Kerr black hole in three dimensions. That 
is the BTZ black hole solution is also a stationary spacetime as that of
Kerr black hole. This implies  that the velocity of photons keeps
probably unchanged 
even for stationary black hole spacetimes.

Before the end of this section, let us consider the propagation of
photons in the magnetically charged black hole backgrounds. In this 
case, the background magnetic field is
\begin{equation}
\bar{F}_{ab}=\frac{Q_m}{4\pi r^2}U_{ab}^{23},
\end{equation}
where $Q_m$ is the magnetic charge of black holes. The equation 
(\ref{eqph5}) describing the propagation of photons  becomes 	
\begin{eqnarray}
\label{eqph6}
k^2 (a \cdot v) &+& \frac{2b(a \cdot v)}{m^2_e}
           [Al^2 +B(m^2 +n^2 -p^2 -q^2)  +Dr^2] \nonumber \\
	&+& \frac{8c}{m^2_e}[A (a \cdot l)(l\cdot v) + 
	B (m \cdot  v)(a \cdot m) +
	B (n\cdot v)(a\cdot n) \nonumber \\
	&& -B(a\cdot p)(v\cdot p) 
        -B(a\cdot q)(v\cdot q) +D(r\cdot a)(r\cdot v)] 
	\nonumber \\
	&-&\frac{1}{m^4_e}\left (\frac{Q_m}{4\pi r^2}\right)^2 
        \left \{32z(a\cdot r)(r \cdot v)+8y[r^2(a\cdot v) 
	+(r \cdot a)(r \cdot v)]\right \}=0.
\end{eqnarray}
It is easy to check that the velocity of radial photons is still $c$, but
the velocities  of orbital photons become
\begin{eqnarray}
&& \left |\frac{k^0}{k^3}\right |_{\rm \phi}
				 =1+\frac{b}{m^2_e}(A+D)-\frac{4y}{m^4_e}
             \left(\frac{Q_m}{4\pi r^2}\right)^2, \\
&& \left |\frac{k^0}{k^3}\right|_{\rm r}=1+\frac{b}{m^2_e}(A+D)
       +\frac{4c}{m^2_e}(A-B)-\frac{4y}{m^4_e}\left (\frac{Q_m}{4\pi
		 r^2}\right)^2 \\ 
&& \left|\frac{k^0}{k^3}\right|_{\rm \theta}=
        1+\frac{b}{m^2_e}(A+D)+\frac{4c}{m^2_e}(B+D)
		  -\frac{16z+8y}{m^4_e}\left(
		  \frac{Q_m}{4\pi r^2}\right)^2 .
\end{eqnarray}
Compared with Eqs.~(\ref{phivel1}), (\ref{rvel1}) and
(\ref{thetavel1}), the contribution from the gravitational background 
keeps unchanged reasonably, the contributions from the magnetic 
field to two physical polarizations interchanges their places.

\section{Photons in radiating topological black hole spacetimes}

In this section we would like to extend the above discussions to
dynamical black hole spacetimes, in which  apparent
horizons and event  horizons do not coincide.   
It is an interesting problem to
see whether or not the ``horizon theorem'' of Shore \cite{Shore} 
holds for the apparent horizons and event horizons in the dynamical
black hole spacetimes. For simplicity, we consider the
topological black hole solutions with a null fluid radiation,
\begin{equation}
R_{\mu\nu}-\frac{1}{2}Rg_{\mu\nu}+\Lambda g_{\mu\nu}= 
8\pi T_{\mu\nu}^{\rm em}+ 8\pi T_{\mu\nu}^{\rm rad}.
\end{equation}
Here  $T_{\mu\nu}^{\rm em}$ is the energy-momentum tensor of 
electromagnetic 
field (\ref{emtensor}) and
 $T^{\rm rad}_{\mu\nu}=\rho \ l_{\mu}l_{\nu}$, is the energy-momentum
 tensor of null fluid with   the 
energy density $\rho$  and the four-velocity $l_{\mu}$.  
  In the advanced  time coordinates, we have \cite{Cai1}
\begin{equation}
\label{dytopo}
ds^2=-f(v,r)dv^2 +2dvdr +r^2[ d \theta ^2 +h^2_k(\theta) d\phi^2 ],
\end{equation}
where 
\begin{equation}
\label{eqf}
f(v,r)=k-\frac{2M(v)}{r}+\frac{Q^2(v)}{4\pi r^2} -\frac{1}{3}\Lambda r^2.
\end{equation}
This is a  generalization  of  the Bonnor-Vaidya metric. The solution 
(\ref{dytopo}) implies that the energy density of null fluid radiation 
is
\begin{equation}
\rho =\frac{\dot{M}(v)}{4\pi r^2} -\frac{Q(v)\dot{Q}(v)}{(4\pi)^2 r^3}, 
\end{equation}
where an overdot represents differentiation with respect to $v$. In the 
solution (\ref{dytopo}) the appropriate basis one-forms are
\begin{equation}
\label{basis}
e^0=\sqrt{f}(dv-f^{-1}dr), \  \ e^1=(\sqrt{f})^{-1}dr, \ \
e^2=rd\theta, \ \ e^3=rh_kd\phi .
\end{equation}
The nonvanishing spin connection one-forms are
\begin{eqnarray}
\label{connection}
&& w^0_{\ 1}=\left( \frac{f'}{2\sqrt{f}}-\frac{\dot
             f}{2f\sqrt{f}}\right)
          e^0 -\frac{\dot f}{2f\sqrt{f}}e^1, \nonumber \\
&& w^1_{\ 2}=-\frac{\sqrt{f}}{r}e^2, \ \ w^1_{\
3}=-\frac{\sqrt{f}}{r}e^3, 
   \ \ w^2_{\ 3}=-\frac{h_{k,\theta}}{rh_k} e^3.
\end{eqnarray}
Using (\ref{basis}) and (\ref{connection}), we obtain the nonvanishing
components of Riemann tensor
\begin{eqnarray}
\label{riecomp2}
&& R_{0101}=\frac{f''}{2}\equiv A,\ \
    R_{0202}=R_{0303}=\frac{f'}{2r}-\frac{\dot f}{2rf}\equiv B, \ \
    R_{0212}=R_{0313}=-\frac{\dot f}{2rf}\equiv E \nonumber \\
&& R_{1212}=R_{1313}=-\frac{f'}{2r}-\frac{\dot f}{2rf}\equiv C, \ \
    R_{2323}=\frac{1}{r^2}\left( k-f\right )\equiv D.
\end{eqnarray}
Now the Riemann tensor can be rewritten as 
\begin{eqnarray}
\label{rieman}
 R_{abcd} &=& AU^{01}_{ab}U^{01}_{cd}+B(U^{02}_{ab}U^{02}_{cd}+
              U^{03}_{ab}U^{03}_{cd})
	  +E(U^{02}_{ab}U^{12}_{cd}+U^{02}_{cd}U^{12}_{ab}
        \nonumber \\
	&+&  U^{03}_{ab}U^{13}_{cd}+ U^{03}_{cd}U^{13}_{ab}) 
	+ C(U^{12}_{ab}U^{12}_{cd}+U^{13}_{ab}U^{13}_{cd})
	+DU^{23}_{ab}U^{23}_{cd},
\end{eqnarray}
and the background electric field is
\begin{equation}
\bar{F}_{ab}=\frac{Q(v)}{4\pi r^2}U^{01}_{ab}.
\end{equation}
In this case, the equation of motion (\ref{eqph5}) for  
photons becomes
\begin{eqnarray}
\label{photon}
k^2 (a\cdot v) &+& \frac{2b}{m^2_e}(a\cdot v)
        \left [ Al^2+B(m^2+n^2)+C(p^2+q^2) 
        +Dr^2 + 2E(m\cdot p +n\cdot q)\right] \nonumber \\
       &+&\frac{8c}{m^2_e}\left [ A(l\cdot v)(a\cdot l )
	 +B(m\cdot v)(m\cdot a) +B(n\cdot v)(n\cdot a)
	 \right.  \nonumber \\
	 && + C(p\cdot v)(p\cdot a) +C(q\cdot v)(q\cdot a)
	 +D(r\cdot v)(r\cdot a) \nonumber \\
 &&\left. +E (m\cdot v)(p\cdot a)+E(p\cdot v)(m\cdot a)
	 +E(q\cdot v)(n\cdot a)
	 +E(n\cdot v)(q\cdot a) \right ] \nonumber \\
	&-&\frac{1}{m^4_e}\left (\frac{Q(v)}{4\pi r^2}\right )^2
	\left \{32z(l\cdot v)(l\cdot a)+8y[l^2(a\cdot v)
	+ (l\cdot v)(l\cdot a)]\right \}=0.
\end{eqnarray}			 
Now it is a position to discuss the propagation of photons in the 
radiating topological black hole spacetimes. Let us consider first 
the case of the radially directed  photons. 

(i). {\it Radial photon motion}. In this case we have  $k^2=k^3=0$.
From Eq.~(\ref{photon}), for the longitudinal polarization $(a\cdot l)$
we have
\begin{eqnarray}
\left \{ k^2 \right. &+& \frac{2b}{m^2_e}[Al^2 +2B(k^0)^2 +2C(k^1)^2 
            + 4E(k^0k^1)] \nonumber \\
	&+& \left. \frac{8c}{m^2_e}Al^2
	-\frac{1}{m^4_e}\left(\frac{Q(v)}{4\pi r^2}
	\right)^2(32z+16y)l^2  \right \}(a\cdot l)=0,
\end{eqnarray}
from  which it follows
\begin{equation}
\label{radr}
\left |\frac{k^0}{k^1}\right|_{\rm r}=1+\frac{2b}{m^2_e}(B+C\pm 2E).
\end{equation}
Here the plus sign corresponds to $k^1>0$ (outgoing photons)
and the minus sign to $k^1<0$ (ingoing photons). For two physical 
polarizations $(a\cdot m)$ and $(a\cdot n)$,
the corresponding equations are
\begin{eqnarray}
\label{radm}
\left \{ k^2 \right. &+& \frac{2b}{m^2_e}[Al^2 +2B(k^0)^2 +2C(k^1)^2 
        + 4E(k^0k^1)] \nonumber \\
	&+& \left. \frac{8c}{m^2_e}[B(k^0)^2 +C(k^1)^2 + 2E(k^0k^1)]
   -\frac{8y}{m^4_e}\left(\frac{Q(v)}{4\pi r^2}\right) ^2
	l^2\right \}(a\cdot m)=0,
\end{eqnarray}
and
\begin{eqnarray}
\label{radn}
\left \{ k^2 \right. &+& \frac{2b}{m^2_e}[Al^2 +2B(k^0)^2 +2C(k^1)^2 
        +4E(k^0k^1)] \nonumber \\
	&+& \left. \frac{8c}{m^2_e}[B(k^0)^2 +C(k^1)^2 +2E(k^0k^1)]
   -\frac{8y}{m^4_e}\left(\frac{Q(v)}{4\pi r^2}\right) ^2
    l^2 \right \}(a\cdot n)=0,
\end{eqnarray}
respectively. From Eqs.~(\ref{radm}) and (\ref{radn}), we obtain 
the velocity of photons
\begin{equation}
\label{radmn}
\left|\frac{k^0}{k^1}\right|_{\theta, \phi}=1+\frac{(2b+4c)}{m^2_e}
        (B+C\pm 2E),
\end{equation}
where the meaning of ``$\pm$'' signs is the same as that
in (\ref{radr}). Here it should be noted that the velocity shift of
photons is changed even  for the radial photons due to the different
motion directions (ingoing  and outgoing photons).

(ii). {\it Orbital photon motion}. In this case we have
$k^1=k^2=0$ for  the $(\phi)$-direction photons. For the
longitudinal polarization
$(a\cdot n)$ it follows from (\ref{photon})
\begin{eqnarray}
\left \{ k^2 \right. &+& \frac{2b}{m^2_e}[A(k^0)^2 +2B(k^0)^2-B(k^3)^2
       +C(k^3)^2 +D(k^3)^2] \nonumber \\
  &+&\left.\frac{8c}{m^2_e}Bn^2 -\frac{8y}{m^4_e}
  \left( \frac{Q(v)}{4\pi r^2}\right )^2 (k^0)^2\right\}
  (a\cdot n) -\frac{8cEn^2}{m^2_e}\left(\frac{k^3}{k^0} \right)
  (a\cdot l) =0.
\end{eqnarray}					  
Note that $n^2=-k^2$. The last term in the above equation can be
neglected because it gives the second correction to the light-cone. Thus
we have 
\begin{equation}
\label{orbn}
\left |\frac{k^0}{k^3}\right|_{\phi}=1+\frac{b}{m^2_e}(A+B+C+D)
         -\frac{4y}{m^4_e}\left (\frac{Q(v)}{4\pi r^2}\right )^2.
\end{equation}
For the transverse polarization $(a\cdot l)$,
\begin{eqnarray}
\left \{ k^2 \right. &+& \frac{2b}{m^2_e}[A(k^0)^2 +2B(k^0)^2-B(k^3)^2
       +C(k^3)^2 +D(k^3)^2] \nonumber \\
    &+&\left. \frac{8c}{m^2_e}[A(k^0)^2+C(k^3)^2]
       -\frac{(32z+16y)}{m^4_e}\left (
  \frac{Q(v)}{4\pi r^2}\right)^2(k^0)^2\right\}(a\cdot l)
  -\frac{8c}{m^2_e}Ek^0k^3 (a\cdot n) \nonumber \\
  &=& 0,
\end{eqnarray}
from which we obtain the velocity of photons
\begin{equation}
\label{orbl}
\left | \frac{k^0}{k^3}\right |_{r}=
     1+\frac{b}{m^2_e}(A+B+C+D)+\frac{4c}{m^2_e}(A+C)-\frac{16z+8y}
	  {m^4_e}   \left(\frac{Q(v)}{4\pi r^2}\right)^2.
\end{equation}
And for the other physical polarization $(a\cdot m)$
\begin{eqnarray}
\left \{ k^2 \right. &+& \frac{2b}{m^2_e}[A(k^0)^2 + 2B(k^0)^2-
         B(k^3)^2 +C(k^3)^2 +D(k^3)^2] \nonumber \\
     &+&\left. \frac{8c}{m^2_e}[B(k^0)^2+D(k^3)^2]
  -\frac{8y}{m^4_e}\left(\frac{Q(v)}{4\pi r^2}\right)^2 
	  (k^0)^2 \right \}(a\cdot m)=0.
\end{eqnarray}
It is straightforward to get the velocity of photons for
$(\theta)$-direction polarization,
\begin{equation}
\label{orbm}
\left |\frac{k^0}{k^3}\right |_{\theta}=
    1+\frac{b}{m^2_e}(A+B+C+D)+\frac{4c}{m^2_e}(B+D)-\frac{4y}{m^4_e}
	 \left(\frac{Q(v)}{4\pi r^2}\right)^2.
\end{equation}
From the components of Riemann  tensor (\ref{riecomp2}), we get
\begin{eqnarray}
\label{eq1}
&& B+C\pm 2E=-\frac{\dot f}{rf}\mp \frac{\dot f}{rf},     \\
\label{eq2}
&& A+B+C+D=\frac{f''}{2}-\frac{\dot f}{rf}+\frac{k-f}{r^2},  \\
\label{eq3}
&& A+C=\frac{f''}{2}-\frac{f'}{2r}-\frac{\dot f}{2rf},   \\
\label{eq4}
&& B+D=\frac{f'}{2r}-\frac{\dot f}{2r f}+\frac{k -f}{r^2}.
\end{eqnarray}

According to the definition of apparent horizon and event horizon
\cite{Haw2}, for static and stationary black holes, the apparent 
horizon  and event horizon coincide with each other. But for the 
radiating black  holes they do no longer coincide. Up to the
$O(L_M,L_Q)$, where $L_M=-\dot M(v)$ and
$L_Q=-\dot Q(v)$, the apparent horizon is  defined as surface whose
expansion $ \Theta = 0$ and the event horizon are null 
surface such that $d\Theta /d v\approx 0$ \cite{Jork}. Generalizing 
the analysis of Koberlein and Mallett \cite{Kob}, for the radiating 
topological black holes (\ref{dytopo}) the apparent horizons can be 
obtained by  roots of  
 the equation $f(v,r)=0$. Inspecting
Eqs.~(\ref{eq1})-(\ref{eq4}), these quantities all diverge at  the 
apparent horizons because $\dot f \ne 0$ at  apparent horizons. This
results in the divergence of velocity shift of photons at the apparent
horizons. It is worth noting that this does not mean that the
velocity of photons becomes infinity at the apparent horizon. In fact,
it reflects that the approximation employed above breaks down in this
case. But an important fact is that the velocity of 
photons is no longer $c$ at the apparent horizon. Another 
quite interesting result is that, from
 (\ref{radr}) and (\ref{radmn}), we find that 
the  velocity is still $c$ for the ingoing photons $(k^1<0)$ of any 
polarization directions; but does not for the outgoing photons 
$(k^1>0)$ and diverges at the apparent horizons.

For the radiating topological black holes (\ref{dytopo}), the event 
horizons, up to $O(L_M,L_Q)$, can also be obtained by the equation 
$f(v,r)=0$, but $M$ and $Q$ in (\ref{eqf}) should be  replaced 
by $M^*$ and $Q^*$ \cite{Kob}.  Here 
\begin{equation}
M^*=M-\frac{L_M}{\kappa}, \  \ Q^*=Q-\frac{L_Q}{\kappa},
\end{equation}
with 
\begin{equation}
\kappa =\frac{M(v)}{r^2_{\rm oah}} -\frac{Q^2(v)}
    {4\pi r^3_{\rm oah}}-\frac{1}{3}\Lambda r_{\rm oah},
\end{equation}
is the surface gravity of outer apparent horizon 
($r_{\rm oah}$) of  black holes.
In this case, these quantities in (\ref{eq1})-(\ref{eq4}) have not 
any strange behavior and finite on the event horizon. Therefore 
the velocity of outgoing radial photons is also no longer  $c$ 
at the event 
horizon. That is, the ``horizon theorem'' of Shore \cite{Shore}
does not hold for the radiating black holes. Inspecting \cite{Shore} reveals 
that the ``horizon theorem'' is derived under the condition that the spacetime 
is stationary. Therefore our result is not inconsistent with the ``horizon
theorem'' because our spacetime is time-dependent. Furthermore, due to
the presence of function $f$ in Eqs,~(\ref{eq1})-(\ref{eq4}), 
the parameter $k$ reflecting the topological structure of spacetimes
enters the expressions of velocity of photons. The velocity 
shift of vacuum polarization photons  becomes  dependent on  
the topological structure of spacetimes.

Now we check the ``polarization sum rule'' in the radiating topological
black holes. With help of (\ref{rieman}) and (\ref{radmn}), for the 
radial motion photons, we have
\begin{eqnarray}
\sum _{\theta, \phi} k^2 &=& -\frac{(8b+16c)}{m^2_e}(B+C\pm 2E)(k^1)^2
            \nonumber\\
           &=&  -\frac{(4b+8c)}{m^2_e}R_{ab}k^ak^b.
\end{eqnarray}
For the orbital photons, we can obtain
\begin{eqnarray}
\sum _{r, \theta}k^2 &=&-\frac{(4b+8c)}{m^2_e}(A+B+C+D)(k^3)^2
          \nonumber \\
	&=& -\frac{(4b+8c)}{m^2_e}R_{ab}k^ak^b,
\end{eqnarray}
where we have assumed $Q(v)=0$. From the above, we can see clearly that
the ``polarization sum rule'' still holds for  the radiating topological 
black  holes.

\section{Conclusion}

The equivalence principle is violated in the quantum field theory in
curved spacetimes, because the  effective action of quantum fields 
contains some interacting terms between quantum fields and spacetime
curvature, which violate the equivalence  principle. Therefore the 
propagation of superluminal photons does not necessarily imply  that 
the causality must be violated  in  curved spacetime. In this work, 
using the one-loop effective action for QED we have investigated, 
respectively, the propagation of photons 
in the static and radiating topological black hole backgrounds.
We have found that in the static topological black hole 
spacetimes, the light cone condition $(k^2=0)$ keeps unchanged for the 
radial motion photons.
That is the velocity of vacuum polarized photons is still $c$ for the 
radial photons. The velocity shifts of orbital photons are same as those 
in the Reissner-Nordstr\"om black hole background. It means that  the
cosmological constant $\Lambda$ and the topological parameter $k$ do 
 not enter the expressions of velocity. This indicates that the propagation
 of vacuum polarized photons is not sensitive to the asymptotic behavior
 and topological structure of backgrounds in this case.
 The light cone condition is also 
unchanged in the massless topological black hole and BTZ black holes,
as in the de Sitter space. Note that the velocity is changed even for 
the radially directed photons due to the stationary feature in the Kerr 
background \cite{Dan2}, and the BTZ black hole spacetime is also a 
stationary one. We conclude that the velocity of photons is not changed 
necessarily even for the stationary black hole spacetimes. It is
important to notice that all  the Riemann tensors have 
the form (\ref{riebtz}) for the massless topological black holes, BTZ
black holes, and (anti-)de Sitter space, in which  the light cone 
condition retains unchanged.

For the propagation of photons 
in the radiating topological black holes, we have  obtained the 
expressions of velocity shifts for radial and orbital photons.
The velocity of photons becomes dependent on the cosmological constant 
and topological structure of backgrounds. We noted that the light 
cone condition of radial photons is changed 
due to the null fluid matter. The velocity of radial photons is no 
longer $c$.   The velocity  will become infinite at the 
apparent horizons, which implies  the breakdown of the 
approximation.  The velocity is also not $c$ at  the event horizons.
Here it should be pointed out that this does not violate the ``horizon
theorem'' of Shore, because the latter is proved for stationary black hole
spacetimes only.  The ``polarization sum rule'' is still valid for the 
radiating topological black holes. 
This is clear because the ``polarization sum rule'' depends on only some 
properties of curvature tensors \cite{Shore},
but the null geodesic plays a crucial role in 
the definition of apparent horizon and event horizon. Therefore it is
unsurprising that the velocity of vacuum polarized photon is no longer
$c$ at the apparent and event horizons. Finally we would like to mention
that the velocity  is still $c$ for the ingoing radial photons. 
We have not yet  understood it.

\section*{Acknowledgments}
This work was supported by the Center for Theoretical Physics of
Seoul National University. The author would like to thank Profs. S.P. Kim,
C.K. Lee, K.S. Soh, and  H.S. Song for a great deal of kind help, C. T. Cho
for useful correspondences, and the referee for helpful comments which helped
him to improve the original version of the paper.


\begin{references}
\bibitem{Haw}S. W. Hawking, Nature 248 (1974) 30; Commun. Math. Phys.
             43 (1975) 199.
\bibitem{Dru}I. T. Drummond and S. J. Hathrell, Phys. Rev. D 22
              (1980) 343.
\bibitem{Ohku}Y. Ohkuwa, Prog. Theor. Phys. 65 (1981) 1058.
\bibitem{Dan1}R. D. Daniels and G. M. Shore, Nucl. Phys. B 425 (1994)
              634.
\bibitem{Alder}S. L. Adler, Ann. Phys. (N.Y.) 67 (1971) 599. 
\bibitem{Dan2}R. D. Daniels and G. M. Shore, Phys. Lett. B 367 (1996)
                  75.
\bibitem{Cho}H. T. Cho, Phys. Rev. D 56 (1997) 6416. 
\bibitem{Shore}G. M. Shore, Nucl. Phys. B 460 (1996) 379.				
\bibitem{Laf}R. Lafrance and R. C. Myers, Phys. Rev. D 51 (1995) 2584.
\bibitem{Schar}K. Scharnhorst, Phys. Lett. B 236 (1990) 354.
\bibitem{Bart}G. Barton, Phys. Lett. B 237 (1990) 559.
\bibitem{Lato}J. I. Latorre, P. Pascual, and R. Tarrach, Nucl. Phys.
               B 437 (1995) 60.
\bibitem{Ben}S. Ben-Menahem, Phys. Lett. B 250 (1990) 133.					
\bibitem{topo}J. P. S. Lemos, Class. Quantum Grav. 12 (1995) 1081; Phys.
        Lett. B 353 (1996) 46; \\
	J. P. S. Lemos and V. T. Zanchin, Phys. Rev. D 54 (1996) 3840.
\bibitem{Huang}C. G. Huang and C. B. Liang, Phys. Lett. A 201 (1995) 27.
\bibitem{Cai1}R. G. Cai and Y. Z. Zhang, Phys. Rev. D 54 (1996) 4891.
\bibitem{Amin}S. Aminneborg, I. Bengtsson, S. Holst, and P. Peldan,
	Class. Quantum Grav. 13 (1996) 2707.
\bibitem{Mann}R. B. Mann, Class. Quantum  Grav. 14 (1997) L 109. 
\bibitem{Bans}M. Ba\~nados, Phys. Rev. D 57 (1998) 1068. 
\bibitem{Brill}D. R. Brill, J. Louko, and P. Peldan, Phys. Rev. D 56
	(1997) 3600.
\bibitem{Van}L. Vanzo, Phys. Rev. D 56  (1997) 6475. 
\bibitem{Cai2}R. G. Cai, J. Y. Ji, and K. S. Soh, ``Topological dilaton
	black holes'' Report gr-qc/9708063 (to be published in Phys. Rev.D).
\bibitem{Maz}F. D. Mazzitelli and F. M. Spedalieri, Phys. Rev. D 52
               (1995) 6694; \\
	M. S. Turner and L. M. Widrow, Phys. Rev. D 37 (1988) 2743. 
 \bibitem{BTZ}M. Ba\~nados, C. Teitelboim, and J. Zanelli, Phys. Rev.
             Lett. 69 (1992) 1849; \\
        M. Ba\~nados, M. Henneaux, C. Teitelboim, and J. Zanelli,
	 Phys. Rev. D 48 (1993) 1506.					
\bibitem{Haw2}S. W. Hawking and G. F. R. Ellis, The Large Scale
               Structure of Space-Time (Cambridge University Press,
	 Cambridge, England, 1973). 
\bibitem{Jork}J. W. Jork, Jr., in Quantum Theory of Gravity: Essays in
               Honor of the Sixtieth Birthday of Bryce S. DeWitt, edited 
	by S. Christensen (Hilger, Bristol, 1984).
\bibitem{Kob}B. D. Koberlein and R. L. Mallett, Phys. Rev. D 49 
             (1994) 5111.

\end{references}
\end{document}